# A microsimulation approach for the impact assessment of a Vehicle-to-Infrastructure based Road Hazard Warning system

**Kallirroi N. Porfyri[1*], Areti Kotsi[1], Evangelos Mitsakis[1]**

[1]Centre for Research and Technology Hellas (CERTH) - Hellenic Institute of Transport (HIT)
E-mail: kporfyri@certh.gr, akotsi@certh.gr, emit@certh.gr
*Corresponding Author

**Abstract**

Cooperative Intelligent Transportation Systems (C-ITS) constitute technologies which enable vehicles to communicate with each other and with road infrastructure. Verification or testing is required for C-ITS applications, in order to assess their impact on traffic operation. In this work, a microscopic traffic simulation approach is used, to evaluate the impact of Vehicle-to-Infrastructure (V2I) technologies in the context of a road traffic accident. Specifically, the methodology is implemented to explicitly models vehicles collisions, Road Hazard Warning (RHW), Emergency Electronic Brake Light (EEBL) warnings and the resulting driver behavior. Moreover, a new gap control mechanism is adopted, to improve safety by advising vehicles in hazard lane to increase their headways with respect to their preceding vehicle, so that they can avoid a collision. Perfect communication links to all vehicles are assumed. The study findings indicate that the proposed V2I hazard warning strategy has a positive impact on traffic flow safety and efficiency.

## *1. Introduction*

Cooperative Intelligent Transportation Systems (C-ITS) constitute an array of connected vehicle technologies anticipated to improve driving experience through the provision of constant and real-time Vehicle-to-Vehicle (V2V) and Vehicle-to-Infrastructure (V2I) services (Baskar, Schutter, Hellendoorn, & Papp, 2011; Fernandes & Nunes, 2012; Kim et al., 2012). The exchange of multiple data collected from field and/ or in-vehicle equipment assists in the deployment of services, which work in a cooperative way, aiming to improve network efficiency, sustainability, safety and environmental impacts mitigation. Concerning road/ traffic safety, the focus is on the reduction of car accidents' risk and on the minimization of unavoidable accidents' resulting damage (Alam, Fernandes, Silva, Khan, & Ferreira, 2015).

Contrary to typical safety warning systems, which utilize information obtained from individual vehicles, V2V and V2I communications apply to a wider range addressing possible implications generated from hazard estimations based on close proximity measures (Milanes et al., 2012).

Technology examples enabling such interactions include Dedicated Short Range Communications (DSRC) based on the IEEE 802.11p standard, and 4G technology using Long Term Evolution (LTE) (Araniti, Campolo, Condoluci, Iera, & Molinaro, 2013; Karagiannis et al., 2011), while interworking between DSRC and cellular network technologies for more efficient Vehicle-to-Everything (V2X) communications has also been proposed (Abboud, Omar, & Zhuang, 2016).

Drivers' behavior with access to V2X technologies is expected to be affected and differentiated from that of conventional vehicles, due to the element of additional information provision. Anticipated benefits are investigated and estimated through simulations, as a more cost effective and less complicated approach compared to field tests. Simulations applicable to such systems cover the aspects of traffic, i.e. vehicle movements, and communication, i.e. message exchanging between vehicles and infrastructure. Under this framework, approaches utilizing the combination of different simulators, in order to address challenges related to components interactions, have been proposed (Queck, Schünemann, Radusch, & Meinel, 2008; Schunemann, Massow, & Radusch, 2008). However, the scope of this work is to examine alterations in driver behavior with V2I integration rather than an in-depth analysis of message communication techniques and requirements.

Taking into account the currently low market penetration rates of V2X communication technologies, microscopic traffic simulation is adopted as an approach enabling accurate modelling of individual vehicles, such as driver acceleration and lane-changing responses to incident and warning messages (Brackstone & McDonald, 1999). Any limitations resulting from assumptions related to zero drivers' mistakes responsible for collisions shall be tackled either by driver behavior modifications (Yang & Peng, 2010) or by computing Surrogate Safety Measures (SSMs) (Caliendo & Guida, 2012; Motro et al., 2016).

The assessment of the impacts of connected vehicle technology-based safety warning systems has been thoroughly studied in the literature. Ye et al. (Ye et al., 2018) develop and evaluate a V2V-based application, performing lane-level hazard prediction, and a corresponding driver's response model. Microscopic traffic simulation results indicate mobility and safety benefits associated to conflicts' decrease, even under low V2V penetration rates. Yeo et al. (Yeo, Shladover, Krishnan, & Skabardonis, 2010) investigate the impact of V2V hazard alert systems on freeway traffic by microscopically modelling equipped and non-equipped driver response to lane-blocking incidents. Simulation outputs show the contribution of V2V hazard alert systems to the mitigation of traffic congestion, as a result of speed reduction and optional lane change messages transmitted to drivers. A hybrid collision warning system for V2V environments, utilizing information collected from vehicle and loop detectors is proposed by Tak et al. (Tak, Woo, & Yeo, 2016). The effectiveness and applicability of the developed system, targeting in overcoming withdraws of typical V2X communication-based collision warning systems (ElBatt, Goel, Holland, Krishnan, & Parikh, 2006; Tak & Yeo, 2013; Wang, Cheng, Lin, Hong, & He, 2008), is tested through the simulation of a vehicle trip. Results lead to the conclusion that the hybrid collision warning system is capable of generating benefits for individual drivers similar to the ones of a V2V system.

In this work the application of a V2I communication-based Road Hazard Warning (RHW) system is supported through LTE technology, in an attempt to avoid any limitations originated from inherent DSRC characteristics. Advantages of LTE include higher market penetration rates, high network capacity, wide cellular coverage range, and technology maturity (Abboud et al., 2016; Araniti et al., 2013). Moreover, perfect communication links to all vehicles that can support communication are assumed.

## *2. Methodology*

In this section the methodological approach adopted to study and evaluate the impact of V2X hazard alert systems in incident situations on road and passenger safety, as well as on traffic efficiency, is elaborated.

### *2.1 Traffic management strategy implementation*

A Long Term Evolution (LTE) data communication network to route V2X data communications is assumed whereas Cooperative Awareness Messages (CAMs) ensure communications among C-ITS enabled vehicles by exchanging continuously data packets with information such as location, speed, identifier, etc. In general, acquired information is integrated to infrastructure through a central server, hereafter denoted as Traffic Control Server (TCS). Subsequently, the TCS generates proposed messages which are taken over by the infrastructure and transmitted to vehicles via V2I communications.

In this work, the road hazard is considered to be a sudden braking (abrupt stopping) of a driver resulting in serious crashes with the following vehicles. Since the location of vehicles can be tracked easily and accurately by the TCS through CAMs, the TCS is able to detect the crash; CAMs indicating abrupt deceleration or sudden stop of the transmission of messages could suggest to the TCS that a crash or a hazardous incident has taken place. When the TCS detects such a conflict, it distributes the corresponding messages to the vehicles in the considered network. The related messages, as well as the drivers' actual responses to a message activation, are described in Table 1.

*Table 1: Drivers' responses to the proposed messages*

| Message | Description | Drivers' response |
|---|---|---|
| Road Hazard Warning (RHW) | Aims to inform drivers in a timely manner of an upcoming hazardous event, as well as how far ahead the hazard is and which lanes are affected. | Drivers on the hazard lane attempt to change lane constantly with the purpose of avoiding a crash. |
| Emergency Electronic Brake Light (EEBL) Warning | Aims to avoid rear-end collisions by warning the drivers that a vehicle in front is suddenly braking hard or crashing. | Drivers on the hazard lane smoothly increase their desired headways to avoid a collision (gap control mechanism). |

| Speed Change Request (SCR) | Aims to inform drivers to change their speed in response to current traffic conditions (crash). | Drivers on the hazard lane and on the adjacent one receive a SCR message to adjust their desired speed. |
|---|---|---|

## 2.2 Traffic management logic for modelling Traffic Control Server and drivers' responses to incidents

This section presents the modelling of drivers' responses to warning messages and the crashing of vehicles using the microscopic traffic simulator SUMO (Simulator of Urban MObility) (Lopez et al., 2018). A major benefit of SUMO is that it is open source, which enables users to develop and integrate new car-following models. Moreover, the simulator includes the Traffic Control Interface tool, shortly known as TraCI, which is a Python API offering users the ability to interact with the running simulation, in order to control the vehicle's parameters, and by extension enabling in case the modelling of drivers' responses, as well as the vehicle collisions.

The default car-following model implemented in SUMO is a variant of the Krauss model, introduced by Stefan Krauss (Krauss, 1998), which is a collision-free model and is based on the same principle as the Gipps car following model (Gipps, 1981). SUMO's inherent default lane change model (i.e. the LC2013 model) also determines the lane changes of a vehicle, which is a sophisticated model that reflects lane change behaviors due to different reasons (e.g. strategic, tactical, mandatory, cooperative, etc.).

In general, the driver's response to the traffic conflict depends on the type of information that is provided by the TCS, as well as on the location of the C-ITS enabled vehicles regarding the road hazard and the other implicated vehicles of the considered network. Taking the above into consideration, we assume that the vehicles are grouped to drive in specific zones; thus, the behavior of a vehicle will be defined with respect to the zone or zones in which it belongs. When the TCS detects that an accident has occurred, it regularly sends RHWs to all vehicles of the predefined influence zones through broadcast communications. The RHW range over which vehicles receive the messages is defined at 500 meters. The communicating vehicles will then adjust their driver behavior by responding to the receiving messages, which correspond to the current traffic conditions. Before proceeding, it is pointed out that we assume perfect communication links to all C-ITS equipped vehicles and that there is no delay between the transmission of messages by the TCS.

In particular, vehicles which travel on the hazard lane and enter the dangerous zone (refer Figure 1) receive a RHW message to adjust their desired speed (SCR), as well as to try to change lane at every time step, if it is feasible; in our model, the vehicles will travel at a lower speed with respect to the speed limit of the freeway after the SCR. Vehicles that drive in the near crash zone (see Figure 1) receive an EEBL warning to inform drivers that a vehicle in front may crash or brake abruptly and therefore the drivers are advised to increase their headways appropriately via a newly developed open-gap function, in order to avoid a collision. Specifically, a gap control mechanism has been recently incorporated into SUMO, allowing the user to impose a smooth adaptation of the vehicle's desired headway. This mechanism aims to facilitate the creation gap between two specific subsequent vehicles and has been modelled to increase the desired time headway of the following vehicle (car-following parameter $tau$), and also determine the minimum space headway that must be maintained between the two vehicles for a predefined duration. The parameters used in the open-gap function are presented in Table 2. Moreover, the TCS decides to which vehicles an EEBL warning will be send, based on the following equation:

$$EEBL_d = RT \times V \times SF \times \frac{V^2}{2d} \quad (1)$$

where, $EEBL_d$ is the distance from the point of initial collision up to which the EEBL warning will be transmitted. $RT$ is the drivers' reaction time and is generally defined as the time that a driver needs to react to avoid an accident after the onset of a threat; the proposed value for reaction time is $RT = 0.9\ s$, which is much lower than the usual reaction time of $2.5\ s$, but is expected for a person reacting to an auditory collision warning (Mohebbi, Gray, & Tan, 2009). $V$ is the speed limit of the road, $SF$ is a safety factor equal to 2 and $d$ is the deceleration ability of vehicles that is set as $4.5\ m/s^2$.

*Table 2: Drivers' responses to the proposed messages*

| Parameter name | Parameter description | Parameter value |
| --- | --- | --- |
| newTimeHeadway [s] | The vehicle's desired time headway will be changed to the given new value with use of the given change rate. | 4 |
| newSpaceHeadway [m] | The vehicle is commanded to keep the increased headway for the given duration once its target value is attained. | 2 |
| duration [s] | The time period in which the time and space headways will be changed to the given new values. | $-1*$ <br> *the largest possible time is set |
| changeRate | The rate at which the new headways' effectiveness is gradually increased. | 0.5 |
| maxDecel [$m/s^2$] | The maximal value for the deceleration employed to establish the desired new headways. | 1.5 |

As regards vehicles that travel in a lane different than the hazard one, they pertain to the safe zone (refer Figure 1). The TCS advises these vehicles to reduce their desired speed with respect to the speed limit of the freeway, whereas they are concurrently restricted from entering the hazard lane. After passing the conflict area, vehicles' speed and lane change operation are no longer under the control of the TCS. Finally, vehicles that have received no information from the TCS are considered to belong in the standard zone (see Figure 1); the behavior of drivers in this state is entirely determined by the default car-following (Krauss) and lane change (LC2013) model implemented in SUMO. The overall traffic management scheme for the modelling of the TCS and the corresponding drivers' responses to messages are described as a flowchart, illustrated in Figure 2.

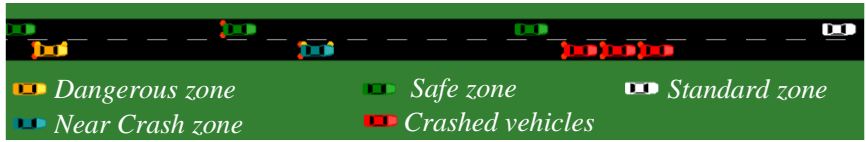

*Figure 1: Representation of the network zones*

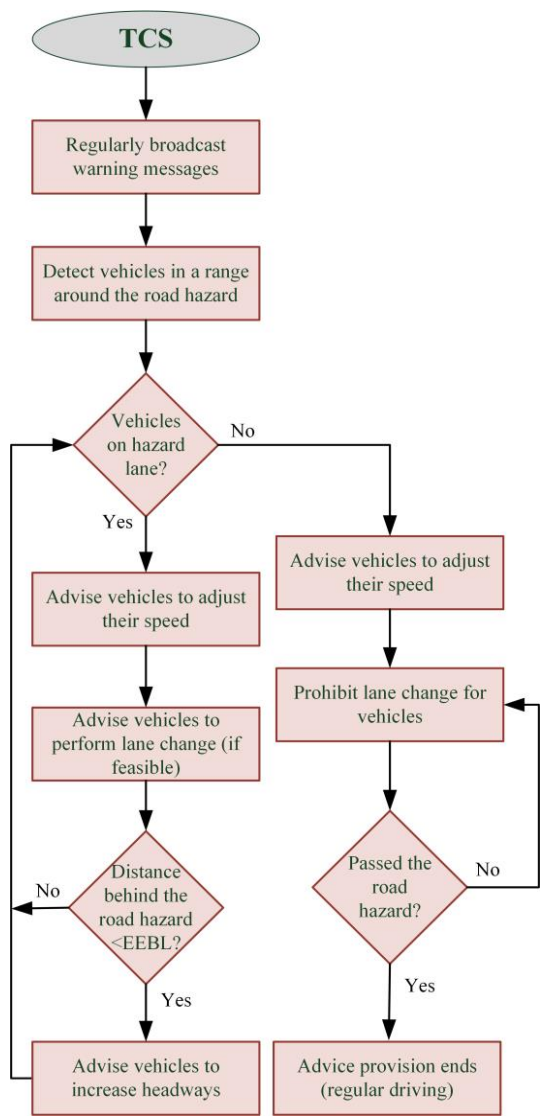

*Figure 2: Flowchart of the proposed traffic management strategy*

*2.3 Surrogate Safety Measures for conflicts points*

This section deals with the description of different traffic conflict indicators for highway safety analysis on a network afflicted by a collision. Specifically, potential SSMs that represent the temporal and spatial proximity characteristics of unsafe interactions and rear-crashes have been used in the simulation-based approach. One of these indicators is the Time to collision (TTC), the most commonly used and well-developed time-based measure. This indicator was initially introduced by Hayard (Hayward, 1972) as an effective measure for rating the severity of traffic conflicts and is defined as the mean time required for two consecutive vehicles to collide if they continue at their current velocity and in the same lane or the same path. In the vast literature, there are different threshold values for the minimum TTC (critical TTC value) required for drivers to safely react, ranging from 1.5 $s$ to 4 $s$ (Dijkstra et al., 2010; Eisele & Frawley, 2004; Essa & Sayed, 2015); the smaller TTC value indicates the higher probability of collision. In this study, a 1.5 $s$ time to collision was adopted as a threshold, which was proposed by the surrogate safety assessment model (SSAM), developed by the Federal Highway Administration (FHWA), to identify a traffic conflict that could lead to a traffic crash (Gettman & Head, 2003; Gettman, Pu, Sayed, & Shelby, 2007). The TTC formulation is as follows:

$$TTC = \frac{D_{LF}}{V_F - V_L} \qquad (2)$$

where, $D_{LF}$ is the distance between the rear bumper of the leading vehicle and the front bumper of the following one (the vehicle under examination); $V_F$ and $V_L$ indicate the speed of the following vehicle and the speed of the leading vehicle, respectively. At this point, it is important to clarify that the TTC is defined for all the follow-lead situations for which the speed of the following vehicle is higher than the speed of the leading one.

A modified SSM derived from TTC is the Time Integrated Time-to-collision (TIT), developed by Minderhoud and Bovy (Minderhoud & Bovy, 2001), and is referred to as the entity of the TTC lower than the threshold. In general, the TIT provides a single value representation of the frequency of crashes and their severity in the area of interest and is also applied to evaluate the risks of traffic conflicts aggregately:

$$TIT = \sum_{i=1}^{N} \int_{0}^{T} max(TTC^* - TTC_i(t), 0) dt \qquad (3)$$

$$\forall 0 \leq TTC_i(t) \leq TTC^*$$

in which $N$ is the total number of vehicles, $TTC^*$ is the specified threshold value of TTC that distinguishes safe from unsafe car-following situations, and $TTC_i(t)$ is the TTC of vehicle $i$ at time $t$.

The concept of safe stopping distance between the leading and following vehicle has also been used, in order to assess the potential risk of collision. Among the most popular safe stopping distance-based indicators for safety estimation is the Deceleration Rate to Avoid Collision (DRAC), which is defined as the minimum rate at which a vehicle must decelerate to avoid a possible traffic conflict. In the context of this study, potential conflict scenarios are considered when the DRAC exceeds a threshold braking value of $3.35 \ m/s^2$ (Archer, 2005). For rear-end interactions, the DRAC indicator can be expressed as:

$$DRAC = \frac{(V_F - V_L)^2}{2D_{LF}} \qquad (4)$$

Besides the above SSMs, several other indicators are described in the literature (Karim, Saifizul, Yamanaka, Sharizli, & Ramli, 2013; Kuang, Qu, & Wang, 2015; Ozbay, Yang, Bartin, & Mudigonda, 2008) that are used for safety assessment of different traffic conflicts through micro-simulation models. However, as the type of vehicle conflict that is modeled in our scenario concerns a rear-end crash that occurs on a freeway stretch and since the aforementioned surrogate indicators are considered to be good measures of the severity of these types of traffic conflicts, TTC, TIT, and DRAC are used as SSMs in our experiment.

## *3. Simulation experiment*

The network used for simulations is a two-lane freeway stretch without any on-ramps and off-ramps; the stretch has a total length of $5 \ km$, with a speed limit of $110 \ km/h$; the simulation time is $1 \ h$ whereas the simulation step is set to $0.1 \ s$. Traffic is flowing at a capacity of 3000 vehicles per hour, constituting of 100% passengers cars. Experiments were conducted for different penetration rates of 0%, 25%, 50%, 75% and 100% C-ITS equipped vehicles to determine their effect on safety and efficiency of the network.

Usually, there are three types of driving: conservative driving, moderate driving and aggressive driving. Therefore, in this study to account for the variation in drivers' behavior and to reflect a more realistic scenario we assume that drivers' behavior follows a normal distribution, which means that most of the drivers have an average aggressiveness in driving (moderate driving behavior) and only a small percentage of drivers would be either very aggressive or conservative in driving. In this respect, the exact degree of aggressiveness for each driver was achieved by setting a normal distribution of sigma, decel, accel and speedFactor parameters in SUMO. Table 3 provides the driver model parameters' values used for our simulation experiment. In case that the entry is not a single number, it has the format normal(<mean>, <std>); [<min>,<max>], specifying a cut off Gaussian distribution.

*Table 3: Driver model parameters values*

| Parameter name | Parameter description | Parameter value |
| --- | --- | --- |
| sigma | The driver imperfection. | $normal(0.2, 0.5)$; $[0.0, 1.0]$ |
| tau [s] | The driver's desired time headway. | 2.0 |
| decel [m/s²] | The maximal deceleration of vehicles. | $normal(3.5, 1.0)$; [2.0, 4.5] |
| accel [m/s²] | The maximal acceleration of vehicles. | $normal(2.0, 1.0)$; [1.0, 3.5] |
| emergencyDecel [m/s²] | The maximal physically deceleration for the vehicles. | 4.5 |
| lcAssertive | Lane-change aggressiveness (willingness to accept lower gaps). | 1.3 |
| actionStepLength [s] | The reaction time. | 0.9 |
| maxSpeed [m/s] | The vehicle's maximum speed. | 30.5 |
| speedFactor | The proportionality factor for the desired speed. | $normal(1.1, 0.2)$; [0.8, 1.2] |

The road hazard that is considered in our experiment is the sudden crash of a vehicle. To simulate this, a vehicle destined for collision is introduced into the network; when this vehicle covers 4 $km$ of the overall length of the freeway stretch, it is made to stop abruptly resulting in serious crashes with the following vehicles. After the onset of the traffic conflict, the resulting behavior of vehicles in the network depends on the predefined by the user values of specific simulation parameters. The parameters that can vary in our experiment are the aforementioned percentage of C-ITS equipped vehicles, as well as the percentage of SCR provided by the TCS. In particular, we assume that the TCS advises vehicles to reduce their desired speed by 15%, 25% and 50% with respect to the speed limit of the freeway. Therefore, this parameterization scheme resulted in a total of 17 different simulation scenarios.

The traffic safety and efficiency of the simulated scenario are evaluated based on the number of crashes that occur, the resulting average network speed and capacity, as well as the three SSMs, i.e. the TTC, the TIT and the DRAC. We assume that events with TTC lower than 1.5 $s$ and DRAC lower than 3.35 $m/s^2$ are safety critical conflicts. The TIT safety indicator is computed for a duration of 15 seconds after the first crash.

## *4. Numerical results*

Table 4 displays key safety metrics for the 17 simulation experiments for several RHW parameterization schemes. Comparison between them reveals that the penetration rate of C-ITS enabled vehicles has a significant impact on the number of crashes, the TIT indicator as well as a positive effect on the capacity of the network. Specifically, the freeway capacity increases when the penetration rate of vehicles that are capable of communication is increased, due to the information that the vehicles receive to change lane markedly prior to the traffic conflict, whereas the number of crashes is significantly reduced with the increase of the penetration rate of C-ITS equipped vehicles. Moreover, a crash reduction is observed as the SCR ratio is decreased. Simulation results also demonstrated that both the increase of equipped vehicles and the lower percentage of SCR result in the reduction of TIT.

*Table 4: Numerical results*

| ID | Penetration rate [%] | Percentage of SCR [%] | TIT | Number of Crashes | Flow [veh/h] |
|---|---|---|---|---|---|
| 1 | 0.00 | - | 128.16 | 5 | 934 |
| 2 |  | 0.50 | 93.75 | 4 | 949 |
| 3 | 0.25 | 0.75 | 94.85 | 5 | 948 |
| 4 |  | 0.85 | 123.34 | 5 | 936 |
| 5 |  | 1.00 | 130.49 | 5 | 933 |
| 6 |  | 0.50 | 90.32 | 3 | 953 |
| 7 | 0.50 | 0.75 | 93.89 | 4 | 950 |
| 8 |  | 0.85 | 104.68 | 5 | 943 |
| 9 |  | 1.00 | 126.71 | 5 | 938 |
| 10 |  | 0.50 | 83.86 | 2 | 958 |
| 11 | 0.75 | 0.75 | 90.42 | 3 | 953 |
| 12 |  | 0.85 | 110.84 | 5 | 942 |
| 13 |  | 1.00 | 118.79 | 5 | 940 |
| 14 |  | 0.50 | 43.16 | 2 | 969 |
| 15 | 1.00 | 0.75 | 59.75 | 2 | 962 |
| 16 |  | 0.85 | 89.4 | 2 | 953 |
| 17 |  | 1.00 | 92.45 | 3 | 951 |

The maximum DRAC and minimum TTC values of the parameterization schemes 1 (0% C-ITS equipped vehicles) and 14 (100% C-ITS equipped vehicles) are illustrated in Figure3; as it can be noticed, in the case of non-equipped vehicles 5 critical situations are indicated with values above the considered thresholds (red dash-dot lines with DRAC= $3.35 \, m/s^2$ and TTC=1.5 s), while in the case of C-ITS equipped vehicles a lower number of critical conflicts are identified (only 2 critical events). Furthermore, the possible critical events that are indicated below the specified thresholds are temporarily shifted and are also lower in number in the case of 100% C-ITS enabled vehicles due to the advice these vehicles receive from the TCS to change lane and increase their headways considerably prior to the crash. Figure 4 depicts the time series mean speed of the traffic flow for non-equipped and 100% C-ITS enabled vehicles; it can be observed that with the operation of RHW the vehicles are moving at higher speeds. Finally, in Figure 5 it can be noticed that the traffic queue observed in the hazard lane for non-equipped vehicles propagates further upstream compared to the case of 100% C-ITS equipped vehicles.

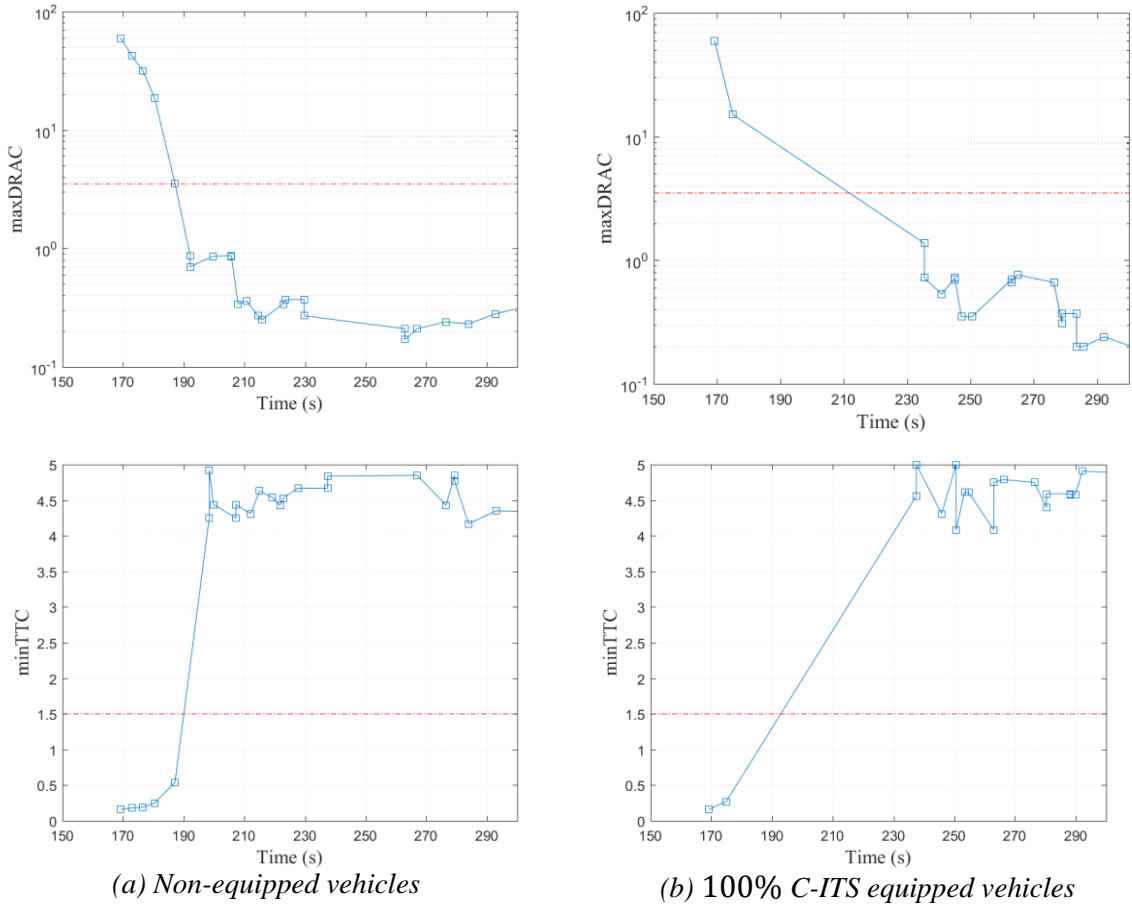

*(a) Non-equipped vehicles*      *(b) 100% C-ITS equipped vehicles*

***Figure 3:*** *DRAC and TTC values over the considered time interval (at the beginning of the traffic conflict for (a) non-equipped and (b) 100% C-ITS equipped vehicles*

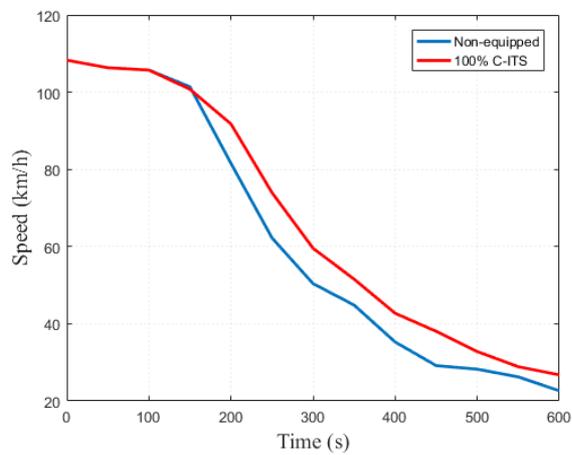

***Figure 4:*** *Average network speed for non-equipped and 100% C-ITS equipped vehicles vehicles*

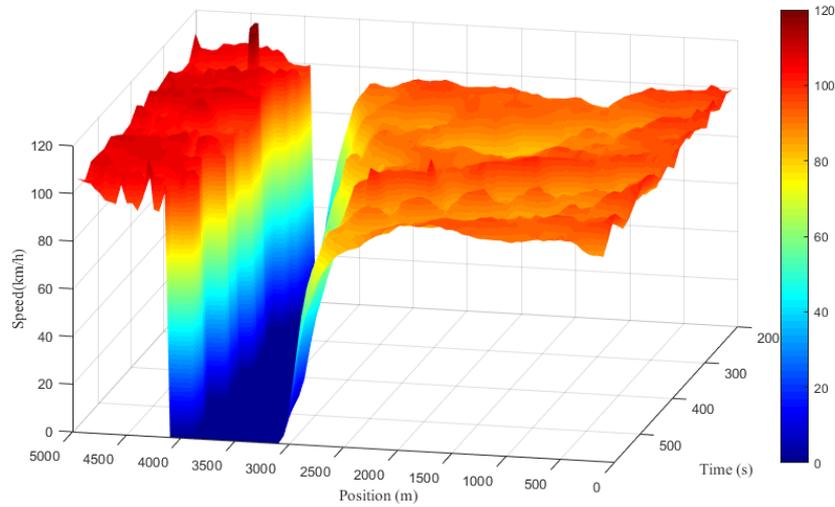

*(a) Non-equipped vehicles*

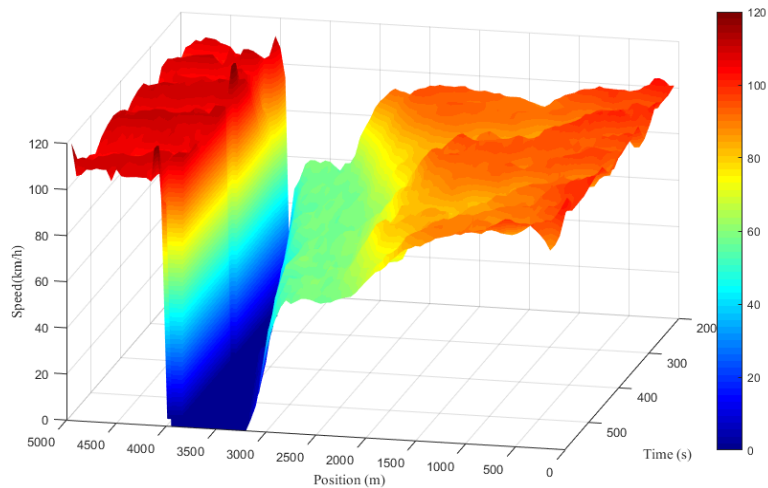

*(b)* 100% *C-ITS equipped vehicles*

**Figure 5:** *Space-time diagram of speed for (a) for non-equipped and (b)* 100% *C-ITS equipped vehicles*

## *5. Conclusions*

In this study, a microsimulation approach was proposed to analyse the effects of a hazard warning system on the behavior and safety of drivers, as well as on the overall traffic efficiency. The implemented methodology explicitly models vehicles collisions, RHW, EEBL warnings and the resulting driver behavior. Moreover, a recently developed gap control mechanism was adopted to improve safety by advising vehicles in the hazard lane to increase their headways appropriately with respect to their preceding vehicle, in order to avoid a collision. Safety was measured using robust SSMs, such as the TTC, the TIT and the DRAC indicators. The study findings demonstrate that the deployment of the proposed V2I hazard warning system has a positive impact on traffic flow safety and efficiency with higher market penetration rates of C-ITS equipped vehicles, particularly if it combines with the reduction of the speed of the vehicles.

## *6. References-Bibliography*